\RequirePackage[l2tabu, orthodox]{nag}
\documentclass[11pt]{article}\usepackage[]{graphicx}\usepackage[]{color}
\makeatletter
\def\maxwidth{ %
  \ifdim\Gin@nat@width>\linewidth
    \linewidth
  \else
    \Gin@nat@width
  \fi
}
\makeatother

\definecolor{fgcolor}{rgb}{0.345, 0.345, 0.345}
\newcommand{\hlnum}[1]{\textcolor[rgb]{0.686,0.059,0.569}{#1}}%
\newcommand{\hlstr}[1]{\textcolor[rgb]{0.192,0.494,0.8}{#1}}%
\newcommand{\hlopt}[1]{\textcolor[rgb]{0,0,0}{#1}}%
\newcommand{\hlstd}[1]{\textcolor[rgb]{0.345,0.345,0.345}{#1}}%
\newcommand{\hlkwb}[1]{\textcolor[rgb]{0.69,0.353,0.396}{#1}}%
\newcommand{\hlkwc}[1]{\textcolor[rgb]{0.333,0.667,0.333}{#1}}%
\newcommand{\hlkwd}[1]{\textcolor[rgb]{0.737,0.353,0.396}{\textbf{#1}}}%

\usepackage{framed}
\makeatletter
\newenvironment{kframe}{%
 \def\at@end@of@kframe{}%
 \ifinner\ifhmode%
  \def\at@end@of@kframe{\end{minipage}}%
  \begin{minipage}{\columnwidth}%
 \fi\fi%
 \def\FrameCommand##1{\hskip\@totalleftmargin \hskip-\fboxsep
 \colorbox{shadecolor}{##1}\hskip-\fboxsep
     \hskip-\linewidth \hskip-\@totalleftmargin \hskip\columnwidth}%
 \MakeFramed {\advance\hsize-\width
   \@totalleftmargin\z@ \linewidth\hsize
   \@setminipage}}%
 {\par\unskip\endMakeFramed%
 \at@end@of@kframe}
\makeatother

\definecolor{shadecolor}{rgb}{.97, .97, .97}
\definecolor{messagecolor}{rgb}{0, 0, 0}
\definecolor{warningcolor}{rgb}{1, 0, 1}
\definecolor{errorcolor}{rgb}{1, 0, 0}
\newenvironment{knitrout}{}{} % an empty environment to be redefined in TeX

\usepackage{alltt}
\usepackage[margin=2cm,footskip=0.5cm]{geometry}

\usepackage[utf8]{inputenc}
\usepackage{lineno}
\usepackage{amsfonts,amsmath,bbm,bm}
\usepackage{float}
\usepackage{etoolbox} %% <- for \cspreto, \csappto
\usepackage{natbib}
\usepackage{thumbpdf}
\usepackage{hyperref}

\newcommand*\linenomathpatch[1]{%
  \cspreto{#1}{\linenomath}%
  \cspreto{#1*}{\linenomath}%
  \csappto{end#1}{\endlinenomath}%
  \csappto{end#1*}{\endlinenomath}%
}

\linenomathpatch{equation}
\linenomathpatch{gather}
\linenomathpatch{multline}
\linenomathpatch{align}
\linenomathpatch{alignat}
\linenomathpatch{flalign}

\newcommand{\R}{\mathbb{R}}

\newcommand{\N}{\mathbb{N}}

\newcommand{\norm}[1]{\left\vert\left\vert #1\right\vert\right\vert}

\renewcommand{\hat}{\widehat}
\newcommand{\Exp}[1]{\hspace{0.7ex} \text{Exp}\left[#1\right]}
\newcommand{\E}[2][]{\hspace{0.7ex} \text{E}_{#1}\left[#2\right]}
\newcommand{\EV}[2][]{\E[1]{2}}
\newcommand{\Var}[2][]{\hspace{0.7ex} \text{Var}_{#1}\left[#2\right]}

\newcommand{\Cor}[2]{\hspace{0.7ex} \text{Cor}\left[#1,~#2\right]}

\newcommand{\comment}[1]{}
\newlength{\solutionHeight}
\newcommand{\solution}[1]{%
	\setbox1=\vbox{#1}%
	\settoheight{\solutionHeight}{\box1}%
	\vspace{\the\solutionHeight}%
	\vspace{\baselineskip}}
\renewcommand{\solution}[1]{~\par {\color{Mulberry} #1} \par \hrulefill~}

\newcommand{\proglang}[1]{\texttt{#1}}
\newcommand{\pkg}[1]{\texttt{#1}}
\newcommand{\code}[1]{#1}

\newcommand{\authornotes}{}
\renewcommand{\maketitle}{{\centering \Large {\makeatletter \textbf{\title} \\ \author \\ \date \par} \vspace{1.5\baselineskip}} {\small \authornotes}}
\renewcommand{\title}{%
Species distribution modelling with spatio-temporal nearest neighbour Gaussian processes
}
\renewcommand{\author}{%
Ethan Lawler$^{1,2}$, Chris Field$^{2}$, Joanna Mills Flemming$^{2}$
}
\renewcommand{\authornotes}{%
\noindent%
1: Author for correspondence - e-mail: lawlerem\makeatletter @\makeatother dal.ca \\
2: Department of Mathematics and Statistics, Dalhousie University, 6316 Coburg Rd, Halifax, NS B3H 1Z9, Canada
}
\renewcommand{\date}{%
}

\IfFileExists{upquote.sty}{\usepackage{upquote}}{}
\begin{document}
%
% \section*{Cover Letter}
%
% We feel that the content of this manuscript could have been submitted as either a Research Article submission or an Application submission, since the contribution is split fairly evenly between the new statistical modelling framework and the implementation in the staRVe package.
% We decided to submit as a Research Article so that we could go into more detail in explaining the statistical model and its use in analyses.
% We have thus tried to keep the style of the paper close to what you would see in a Research Article, but to help advertise the package there are a few small sections that are closer to an Application style.
%
%
%
% \newpage

\maketitle

% \linenumbers

% \title
%
% \author

\section*{Abstract}

\begin{enumerate}
  \item{Spatio-temporal datasets that are difficult to analyze are common in ecological surveys. There are software packages available to analyze these datasets, but many of them require advanced coding skills. There is a growing need for easy to use packages that researchers can use to analyze common ecological datasets.}
  \item{We develop a particular generalized linear mixed model for spatio-temporal point-referenced data that is flexible enough to accommodate data from most ecological surveys while being structured enough to facilitate analyses without advanced coding. Our implementation in the {\bfseries staRVe} package uses a computationally efficient version of a nearest neighbour Gaussian process enabling analysis of relatively large datasets.}
  \item{A brief simulation study shows our model produces accurate predictions and forecasts, while a tutorial analysis of a Carolina wren survey suggests a recommended workflow for analyses. We also analyze a more complicated scientific survey of haddock to showcase the capabilities of our model.}
  \item{Our model and package are tools that can easily be added to researchers' workflow to help make sense of data from ecological surveys. We emphasize the ability of our model to create useful visualisations of data which can then lead to identification of important trends in species distributions.}
\end{enumerate}

\section*{Keywords}

hierarchical model, generalized linear mixed model, nearest neighbours Gaussian process, spatio-temporal analysis, species distribution model

\section{Introduction}

Spatio-temporal datasets that are difficult to analyze are commond in ecological surveys.
Such surveys typically record either the count, weight, or simply presence or absence of a species observed within a small spatial transect.
To track the health of the population through time surveys are typically repeated at the same time every year.
This leads to data that represent points in continuous space and discrete time.
In some surveys the same spatial transects are sampled every year, although others such as marine surveys use a stratified random sampling design which leads to different transects being sample every year.
Repeated transect designs are much easier to analyze and associated spatially explicit statistical models have a relatively long history in the literature.
The stratified random sampling designs have traditionally been modelled on aggregate by discarding the spatial information, with spatially explicit models appearing only recently.
Statistical models for both types of sampling design, however, rely in some form on Gaussian processes.

Gaussian processes underlie nearly all statistical modelling of spatial data so we briefly review the recent literature and relevant software.
A Gaussian process is a (possibly infinite) collection of random variables where any finite subset has a multivariate normal distribution, and can be fully specified by a mean function and covariance function.
Fitting models with a Gaussian process is computationally challenging, and recent applied interest in analyzing large spatial or spatio-temporal datasets has led to theoretical innovations making these models computationally tractable.
Many of these innovations introduce different ways to approximate a Gaussian process -- either by approximating the process itself or by imposing a low-rank condition on the covariance matrix.
Examples of methods that approximate the process are the stochastic partial differential equation (SPDE) approach \citep{Lindgren2011::spdeGMRF} and the nearest-neighbour Gaussian process \citep{Datta2016::nearestNeighbourGP}.
Low-rank methods include fixed-ranked kriging \citep{Cressie::FixedRankKriging} and the predictive process \citep{Banerjee2008::predictiveProcess}.
An enjoyable case-study competition between a wide variety of these methods is described in \citet{Heaton2018::LargeSpatialCaseStudyCompetition}.

Some of the aforementioned examples have existing and well-developed \proglang{R} packages.
However due to their flexibility they often require expert knowledge of \proglang{R} to correctly code the desired model.
Perhaps the most widely used package for spatial ecology is \pkg{R-INLA} \citep{Lindgren::R-INLA}, which implements the SPDE approach mentioned earlier.
Another popular package in spatial ecology is Template Model Builder, or \pkg{TMB} \citep{Kristensen2016::TMB}.
While \pkg{TMB} is not connected with any particular Gaussian process approximation, it takes advantage of the Laplace approximation and automatic differentiation for fast optimization of a user-coded likelihood function, and was developed with spatial applications in mind.

The \pkg{R-INLA} and \pkg{TMB} packages are quite flexible, but require the user to spend a significant amount of time learning how to code the model or likelihood function.
What is currently missing is a suite of more specialized \proglang{R} packages that are easy to use in specific domains.
A few of these packages already exist, such as \pkg{LatticeKrig} \citep{Nychka::LatticeKrig} for purely spatial data or \pkg{SpatioTemporal} \citep{Lindstrom::SpatioTemporalPackage} for fitting Gaussian spatio-temporal processes using basis functions.
For more case-specific packages that support spatial analyses we have \pkg{Hmsc} \citep{Tikhonov2020::HmscR} for multi-species community data and the \pkg{VAST} package \citep{Thorson2017::VAST} which is popular in fisheries research and supports spatio-temporal data.

To the above literature we add a new dynamic spatio-temporal model with its own flavour of modelling.
Our model uses a new spatio-temporal version of the nearest neighbour Gaussian process, and is implemented using the \pkg{TMB} package to provide computationally fast inference, simulations, and predictions.
This framework is specific enough to admit a user-friendly interface with little advanced programming knowledge required.
At the same time many applied examples fit the continuous-space discrete-time data description giving the package wide applicability.
Examples include daily ozone levels, yearly cancer incidence levels, and the ecological examples we explore in this paper.
Our initial motivating domain in fisheries research also gives our package its acronym: {\bfseries S}patio-{\bfseries T}emporal {\bfseries A}nalysis of {\bfseries R}esarch {\bfseries VE}ssel data, or \pkg{staRVe}.

\section{Methods and Materials}

\subsection{Modelling Framework}
\label{ssec::modelDescription}

We assume that we have univariate spatio-temporally referenced point data in continuous space $\mathbf{s}\in\R^{d}$ (or a connected subset of $\R^{d}$) and discrete time $t\in \N$.

Our model can be written in the familiar generalized linear mixed model framework:
\begin{align*}
    \mathbf{Y} ~\vert~ \mu_{y} &\sim f_{\theta}\left(y; \mu_{y}\right) \\
        \mu_{y} ~\vert~\mathbf{w} &= g^{-1}\left(\mathbf{X}\bm{\beta}+\mathbf{Z}\mathbf{w}\right)
\end{align*}
where $\mathbf{Y}$ is the response variable, $f_{\theta}\left(y; \mu_{y}\right)$ is a response distribution depending on parameters $\theta$ with mean $\mu_{y}$, $g^{-1}$ is an inverse link function, $\mathbf{X}$ is a design matrix for fixed effects/covariates with coefficients $\bm{\beta}$, and $\mathbf{Z}$ is a design matrix for random effects given by $\mathbf{w}$.
To remove identifiability issues between an intercept term in $\mathbf{X}$ and the mean of the random effects $\mathbf{w}$, we exclude the intercept term in $\mathbf{X}$ \citep{Ogle2020::HierarchicalIdentifiability}.
For notational simplicity, we replace the design matrix $\mathbf{Z}$ with indices specifying the time point and location, so that an observation $Y_{i,t}\left(\mathbf{s}\right)$ at time point $t$ and location $\mathbf{s}$ is dependent on the random effect $\mathbf{w}_{t}\left(\mathbf{s}\right)$.
With this change we write the spatio-temporally referenced model as
  \begin{align*}
    \mathbf{Y}_{i,t}\left(\mathbf{s}\right) ~\vert~ \mu_{t}\left(\mathbf{s}\right)
      &\sim f_{\theta}\left(y_{i,t}\left(\mathbf{s}\right);~\mu_{t}\left(\mathbf{s}\right) \right) \\
    \mu_{t}\left(\mathbf{s}\right) ~\vert~ \mathbf{w}_{t}\left(\mathbf{s}\right)
      &= g^{-1}\left(\mathbf{X}_{i}\bm{\beta} + \mathbf{w}_{t}\left(\mathbf{s}\right)\right)
  \end{align*}
The interesting modelling goes into specifying the distribution for the random effects $\mathbf{w}_{t}\left(\mathbf{s}\right)$.

\paragraph{The spatio-temporal random effects}

We model our spatio-temporal random effects as a time series of spatial fields, or a spatial field that evolves over time, using a two-level hierarchical specification.
The 'temporal' level in the hierarchy models the global temporal evolution of the random effects.
The 'spatial' level in the hierarchy models the spatial variation in the random effects, or more precisely, it models the spatial variation in how each location evolves from one time to the next.

The 'temporal' level is specified as a simple autoregressive order 1, or AR(1), process.
Specifying a set of temporal random effects $\bm{\epsilon}$, with a random effect $\epsilon_{t}$ for each time point in the model, we have
\[
  \bm{\epsilon} \sim
    \text{N}\left(\mu,~\frac{\sigma^{2}}{1-\phi^{2}}\cdot \Sigma\right)
    ~~\text{where}~~\Sigma_{ij} = \phi^{\norm{i-j}}.
\]
where $\mu$ is the mean of the process, $\sigma^{2}$ is the one-step-ahead variance $\Var{\epsilon_{t+1}~\vert~\epsilon_{t}}$, and $\phi$ is the correlation between adjacent times $\Cor{\epsilon_{t+1}}{\epsilon_{t}}$.

Moving to the 'spatial' level of the hierarchy, the Gaussian process (GP) is a ubiqitous choice for any spatial process due to its attractive stochastic properties
A Gaussian process is a generalization of a multivariate normal random variable to infinite dimension, where any finite subset has a multivariate normal distribution.
However, the computation cost in evaluating the likelihood for a Gaussian process is prohibitive even for modestly sized datasets.
We use a nearest-neighbour Gaussian process (NNGP) as a computationally efficient alternative \citep{Datta2016::nearestNeighbourGP}. More detail than is given here can be found in appendix \ref{app::likelihoodDetails}.
A GP is described completely by a mean function
	\(
		\mathbf{m}\left(\mathbf{s}\right): \R^d \to \R
	\)
  that gives the mean at each location $\mathbf{s}$, and a covariance function
	\(
		\mathcal{C}\left(\mathbf{s}_{1},\mathbf{s}_{2}\right): \left(\R^{d},\R^{d}\right) \to \R^{+}.
	\)
The covariance function encodes Tobler's first law of geography that "everything is related to everything else, but near things are more related than distant things" \citep{Tobler1970::movieSimulationUrbanGrowth}.
Additionally the NNGP requires a directed acyclic graph to be specified on a set of random effect nodes, acting much like the mesh used in the \pkg{R-INLA} package.

We have a different spatial process at each time point, so we index these random effects by both time and space:
  \[
    \mathbf{W}_{t}\left(\mathcal{S}\right) \sim
      \text{NNGP}\left(\mathbf{m}_{t}\left(\mathbf{s}\right),~\mathcal{C}\left(\mathbf{s}_{1},\mathbf{s}_{2}\right)\right).
  \]
The 'temporal' level in our hierarchy is used to inform the mean function at each time point
  \[
    \mathbf{m}_{t}\left(\mathbf{s}\right) = \phi\cdot\left(\mathbf{w}_{t-1}\left(\mathbf{s}\right)-\epsilon_{t-1}\right) + \epsilon_{t}.
  \]
By including the term $\phi\cdot\left(\mathbf{w}_{t-1}\left(\mathbf{s}\right)-\epsilon_{t-1}\right)$ we ensure that the temporal evolution at any specific location follows an AR(1) process.
The time series at each location are spatially correlated, the strength of which is described by the covariance function.

The Mat\'ern covariance function is the most widely used covariance function in the literature, containing many special cases such as the exponential.
Here we present only the exponential covariance function for simplicity.
  \[
    \mathcal{C}\left(\mathbf{s}_{1},\mathbf{s}_{2}\right)
      = \tau^{2}\rho
        \Exp{-\frac{ \norm{\mathbf{s}_{1}-\mathbf{s}_{2}} }{ \bar{d} }
          \cdot \frac{1}{\rho}}.
  \]
We estimate the spatial standard deviation parameter $\tau$ and spatial range parameter $\rho$.
The average distance $\bar{d}$ between 'connected' random effect nodes in the directed acyclic graph is included to standardise parameter estimates for datasets with different spatial scales (technically, it is the spatial scale of the random effect nodes that is standardised).

The spatial range parameter $\rho$ is nominally the distance at which locations are approximately (spatially) uncorrelated.
The range parameter is weakly estimable but strictly speaking it is unidentifiable with a mostly flat profile likelihood, and does not affect inference of other model parameters nor prediction of the random effects \citep{zhang2004::InconsistentEstimation}.
Since estimating the parameter is fairly difficult a common approach is to fix the value of the $\rho$ to be some large value, which can in some cases provide predictions that are nearly as good as they would be if the true value of $\rho$ were known \citep{Kaufman2013::RangePredictionGeostatistics}.
When possible though, we still recommend trying to estimate $\rho$ via maximum likelihood.

\paragraph{The log-likelihood function}

Writing the complete model, we have
\begin{align*}
	\mathbf{Y}_{i,t}\left(\mathbf{s}\right) ~\vert~ \mu_{i,t}\left(\mathbf{s}\right)
    &\sim f_{\theta}\left(y_{i,t}\left(\mathbf{s}\right);~\mu_{i,t}\left(\mathbf{s}\right) \right) \\
  \mu_{i,t}\left(\mathbf{s}\right) ~\vert~ \mathbf{w}_{t}\left(\mathbf{s}\right)
    &= g^{-1}\left(\mathbf{X}_{i}\bm{\beta} + \mathbf{w}_{t}\left(\mathbf{s}\right)\right) \\
	\mathbf{W}_{t}\left(\mathcal{S}\right) ~\bigr\vert~ \left\{\mathbf{w}_{t-1}\left(\mathcal{S}\right),~\bm{\epsilon}\right\}
    &\sim \text{NNGP}\left(
	  			\phi\cdot\left(\mathbf{w}_{t-1}\left(\mathbf{s}\right)-\epsilon_{t-1}\right)+\epsilon_{t},~
	  			\mathcal{C}\left(\mathbf{s}_{1},\mathbf{s}_{2}\right)\right) \\
  \bm{\epsilon} &\sim
    \text{N}\left(\mu,~\frac{\sigma^{2}}{1-\phi^{2}}\cdot \Sigma\right)
    ~~\text{where}~~\Sigma_{ij} = \phi^{\norm{i-j}}.
\end{align*}
The mean for the initial spatial field is spatially constant and equal to $\epsilon_{1}$.

We proceed with maximum likelihood inference and write the joint log-likelihood function as
	\(
		\ell_{J}\left(\Theta\right) = \ell_{y}\left(\Theta\right) + \ell_{w}\left(\Theta\right)
	\)
	where $\ell_{y}\left(\Theta\right)$ is the likelihood contribution from the response distribution and $\ell_{w}\left(\Theta\right)$ is the likelihood contribution from the random effects in both the 'temporal' and 'spatial' levels of the hierarchy.
We integrate out the random effects and use the restricted or marginal log-likelihood function
	\[
		\ell\left(\Theta\right)
			= \int \ell_{J}\left(\Theta\right) d\mathbf{W}
			= \int \ell_{y}\left(\Theta\right) + \ell_{w}\left(\Theta\right) d\mathbf{W}.
	\]
The above integral is computed using the Laplace approximation as implemented in the \proglang{R} package \pkg{TMB} \citep{Kristensen2016::TMB}.
We obtain maximum likelihood estimates $\hat{\Theta}$ (with standard errors) for the parameters $\Theta = \left\{\theta,~\bm{\beta},~\mu,~\phi,~\sigma^{2},~\tau\right\}$.
Predictions (with standard errors) for the random effects $\mathbf{W}_{t}\left(\mathbf{s}\right)$ and $\bm{\epsilon}$ are then obtained by maximizing the joint likelihood $\ell_{J}\left(\hat{\Theta}\right)$ with respect to the random effects.

\paragraph{Predictions}

A common task in spatial and spatio-temporal statistics is to predict the random effects and response mean at unobserved sites, either locations which have not been observed or forecasting at future times.
We obtain predictions in three steps: first the random effects, then the linear predictor including the covariates, and finally the response mean after applying the link function.
These predictions are easily obtained from our model coupled with the functionality in the \pkg{TMB}.

To our joint likelihood $\ell_{J}\left(\Theta\right)$ we add another contribution $\ell_{p}\left(\Theta\right)$, the conditional distribution of the random effects at the prediction locations given the random effects in $\ell_{J}\left(\Theta\right)$.
The predicted value for $W_{t}\left(\mathbf{s}\right)$ is then taken to be the mode of $\ell_{J}\left(\hat{\Theta}\right)+\ell_{p}\left(\hat{\Theta}\right)$ as a function of $w_{t}\left(\mathbf{s}\right)$, and the standard errors are computed from the hessian evaluated at the mode.
This is essentially a plug-in estimate using the maximum likelihood estimates $\hat{\Theta}$, but since we take the predictions directly from the likelihood function all uncertainty in parameter estimation is propogated to the predictions of the random effects.

With a new set of covariates $\mathbf{X}$, regression coefficient estimates, and predicted random effects we obtain the linear predictor
	\[
	  \mathbf{X}\hat{\bm{\beta}}+\hat{\mathbf{w}}_{t}\left(\mathbf{s}\right),
	\]
Finally, the predicted response mean is found by applying a second-order Taylor approximation to
 	\(
		g^{-1}\left[
			\mathbf{X}\hat{\bm{\beta}}+\hat{\mathbf{w}}_{t}\left(\mathbf{s}\right)
		\right].
	\)

\subsection{Simulations}

To develop an understanding of our model framework we visualise data simulated from different scenarios within it.
By visualising these simulations we help develop an intuition for parameter interpretation and the performance of the package.
Our scenarios are
\begin{enumerate}
  \item{spatial simulations with no observation error}
  \item{spatio-temporal simulations with Gaussian response}
  \item{spatio-temporal simulations with Poisson response}
\end{enumerate}
We take the simulation design used by \citet{zhang2004::InconsistentEstimation} and adapt it to visualise our model in the spatio-temporal setting.
We simulate random effects on a grid covering the unit square, with points at $(0.1*x,0.1*y)$ for $x,y\in 0,~1,~...,~10$ and $(0.1*x+0.05,0.1*y+0.05)$ for $x,y\in 0,~1,~...,~9$ for a total of 221 locations.
We also simulate random effects on a transect covering $(0.387,0.01*y)$ for $y\in 0,~...,~100$.
For our two spatio-temporal scenarios we simulate data for ten years.
We then apply the inverse link function to the simulated random effects to get the expected value for each data point, then simulated data from the corresponding response distribution.
The data for scenario 1 are taken to be the simulated random effects, which corresponds to a scenario of no observation error.

We then fit a model to the simulated data on the grid, and use the fitted model to produce predictions for the transect.
In the two spatio-temporal scenarios we fit the model using data from years one through eight and then use the fitted model to forecast predictions into years nine and ten.
For all scenarios we simulate with the spatial range $\rho=0.3$ and estimate it via maximum likelihood.

\paragraph{Spatial simulations with no observation error}

We simulate a realization of the Gaussian random effects, without observation error, and for a single year.
Since $\mu$ only shifts the data up or down we let $\mu=0$.
The temporal parameters $\phi$ and $\sigma$ do not enter the model since there is only a single year of observations.
The only remaining parameter in this scenario is the spatial standard deviation $\tau$.
We simulate the random effects using the value $\tau=1$.
To see the effect of $\tau$ on the model we fix the estimated value of $\tau$ at three pre-selected values $\tau=0.5$, 1, and 2.
We then use the fitted model to produce predictions for the transect and compare the point predictions and standard errors to see the effects of $\tau$.

\paragraph{Spatio-temporal simulations with Gaussian response}

We simulate a realization of a dataset using a Gaussian response distribution with eight years of data to fit the model, and an additional two years to forecast predictions.
We set the temporal autocorrelation parameter $\phi=0.5$ and response variance $\sigma_{\text{obs}}=2$.
The mean $\mu$ is still equal at zero.
We simulate data under four different spatio-temporal characteristics, the combinations of low or high spatial variance ($\tau=1$ and $\tau=3$) and low or high temporal variance ($\sigma=5$ and $\sigma=15$).
Unlike in the previous section we fit all models using maximum likelihood, and plot the predicted mean and 95\% confidence band only for ML parameter estimates.
We predict and plot the random effects instead of the data to remove the observation noise when plotting the simulated data.

\paragraph{Spatio-temporal simulations with Poisson response}

We replace the Gaussian response distribution of the previous simulation settings with a Poisson distribution.
When plotting we convert the random effects to the response scale so the simulations and predictions shown represent the spatio-temporally varying rate parameter of a Poisson distribution.
We set the autocorrelation parmaeter $\phi=0.5$ and spatial standard deviation $\sigma=0.5$.
We then simulate data under four different combinations: low or high spatial variance ($\tau=0.05$ and $\tau=0.2$) and low or high global mean ($\mu=\log{8}$ and $\mu=\log{15}$).

\paragraph{Interpreting the plots}

We plot only the transect instead of the whole domain, which allows us to include multiple years of the same transect in a single plot.
The transect is labeled from south (S) to north (N) corresponding to the points $(0.387,0)$ and $(0.387,1)$ respectively.
When plotting multiple years of the same transect, we paste the transect corresponding to year $t$ at the end of the transect for year $t-1$.
The years are labelled and separated into vertical bands by a dashed line.
The spatial variability is the amount of variation the plot exhibits within each vertical band (i.e.,~the south-to-north transect within ad particular year).
The temporal variability is the amount of variation the plot exhibits from one vertical band to the next (i.e.,~the evolution of the transect from one year to the next).

\subsection{Carolina Wren Survey}

We analyze a dataset containing the number of sightings of Carolina wren {\itshape Thryothorus ludovicianus} recorded once a year along spatial transects in Missouri, U.S.A taken from the package \pkg{STRbook} \citep{Zammit::spatiotemporalWithRpkg}, which in turn was modified from \citet{Pardieck::birdSurveyData}.
This analysis is meant to demonstrate the oft-unreported typical steps -- from model fitting and checking to predictions and interpretation -- and to highlight some of the surrounding \proglang{R} packages that our implementation is enhanced by.

The Carolina wren dataset is stored using the ``simple features" open standard for spatial vector data as implemented in the \proglang{R} package \pkg{sf} \citep{Pebesma::sf}.
We could not find the coordinate reference system used for this data, so for demonstrative purposes we assume it is WGS84.
The dataset has 783 observations across 21 years, ranging from 1994 to 2014.
The same locations are surveyed every year, however there are some missing observations which have been removed from the dataset.
Without these missing data, there is a median 36 observations per year, with as few as 26 and as many as 45.

Our initial model for this dataset uses a Poisson distribution with the canonical log link function.
The convergence message is ``relative convergence", so we are confident that we have found a local maximum of the log-likelihood function.
With ecological count data it is usually a good idea to check for over- or under-dispersion of the data relative to the fitted model.
While over- or under-dispersion can be caused by any number of model mis-specifications, the easiest one to check is for mis-specification of the response distribution.
We use quantile residuals provided by the \pkg{DHARMa} package, which uses simulations to construct a parametric bootstrap estimator of the cumulative distribution function (CDF) for each data point \citep{Hartig2020::DHARMa}.
If the model and response distribution are correct then the residuals from the \pkg{DHARMa} package will be uniformly distributed on the interval [0,1].

For our analysis we simulate 100 set of observations from the fitted model to construct the bootstrap CDF.
We then simulate one set of observations each from a model with a Poisson response distribution, an over-dispersed negative binomial response distribution, and an under-dispersed Conway-Maxwell-Poisson response distribution.
The Conway-Maxwell-Poisson distribution can be seen as a generalized version of the Poisson distribution that can exhibit both under- and over-dispersion and has been used in a variety of applications \citep{Sellers2012::compois}.
To ensure we are checking only the response distribution we simulate new observations conditional on the fitted random effects and parameters, so that all of the simulated datasets share the same spatio-temporal pattern as the Carolina wren dataset.
Quantile residuals for the Carolina wren dataset and each of the different response distributions are computed relative to the Poisson model, i.e.~using the bootstrap CDF from the fitted model.
Through this procedure we have four sets of residuals: one set where we know the model is correct (Poisson simulation), one set where we know the data is over-dispersed relative to the model (negative binomial simulation), one set where we know the data is under-dispersed relative to the model (Conway-Maxwell-Poisson simulation), and the final set of residuals coming from the real data set.
We then compare the residuals from the Carolina wren dataset to the residuals coming from the simulated datasets to determine whether the data exhibit under- or over-dispersion relative to the fitted model using the Poisson distribution.

After checking the fitted model we inspect the parameter estimates to get a general sense of spatio-temporal behaviour of the random effects and which covariates have significant effects.
With those general impressions in mind we then create a spatio-temporal map of the response mean as an important visualisation of the data and fitted model.
We produce a smooth map by predicting the response mean on a raster using the procedure detailed in Section \ref{ssec::modelDescription}, and using the raster data format supplied by the \pkg{raster} package \citep{Hijmans::raster}.
The final graphic is produced using the raster predictions as a data layer in the \pkg{tmap} package \citep{Tennekes::tmap}, though the raster could also be exported to an external GIS program.

\paragraph{Code Excerpts}

To help make these methods available for use we have developed an easy-to-use \proglang{R} package implementing our model framework.
We show the relevant code for fitting the Conway-Maxwell-Poisson model to the Carolina wren dataset, and then using the fitted model to predict mean intensity.
The code to fit a model should be familiar to those who have fit a generalized linear model in \proglang{R} with only a model formula, a dataset, and a response distribution needed.
We use the option `fit=T' since we do not need to customize the model past the default settings.

\begin{knitrout}
\definecolor{shadecolor}{rgb}{0.969, 0.969, 0.969}\color{fgcolor}\begin{kframe}
\begin{alltt}
\hlstd{bird_compois}\hlkwb{<-} \hlkwd{prepare_staRVe_model}\hlstd{(}
  \hlstd{cnt}\hlopt{~}\hlkwd{time}\hlstd{(year),}
  \hlstd{bird_survey,}
  \hlkwc{distribution}\hlstd{=}\hlstr{"compois"}\hlstd{,}
  \hlkwc{fit}\hlstd{=T}
\hlstd{)}
\end{alltt}
\end{kframe}
\end{knitrout}

We then use the fitted `bird\_compois' model to predict the mean intensity throughout the state of Missouri.
We want a smooth map so we first create a template raster to tell where predictions should be made.
The first few lines of code below create this raster, and mask the portions of the raster that are outside of Missouri state lines.
The last line of code then takes the fitted model and the template raster and gives a list of rasters containing the predictions for each year.
For this analysis we use the default behaviour of the `staRVe\_predict' function which gives predictions for each year included in the dataset used to fit the model.

\begin{knitrout}
\definecolor{shadecolor}{rgb}{0.969, 0.969, 0.969}\color{fgcolor}\begin{kframe}
\begin{alltt}
\hlstd{missouri}\hlkwb{<-} \hlkwd{subset}\hlstd{(}\hlkwd{ne_states}\hlstd{(}\hlkwc{iso_a2} \hlstd{=} \hlstr{"US"}\hlstd{,}\hlkwc{returnclass}\hlstd{=}\hlstr{"sf"}\hlstd{),}
        \hlstd{name} \hlopt{==} \hlstr{"Missouri"}\hlstd{,}
        \hlkwc{select} \hlstd{=} \hlstr{"name"}\hlstd{)}
\hlkwd{st_crs}\hlstd{(missouri)}\hlkwb{<-} \hlnum{4326}
\hlstd{raster_to_pred}\hlkwb{<-} \hlkwd{rasterize}\hlstd{(missouri,}\hlkwd{raster}\hlstd{(missouri,}\hlkwc{nrow}\hlstd{=}\hlnum{20}\hlstd{,}\hlkwc{ncol}\hlstd{=}\hlnum{20}\hlstd{),}\hlkwc{getCover}\hlstd{=T)}
\hlstd{raster_to_pred[raster_to_pred} \hlopt{==} \hlnum{0}\hlstd{]}\hlkwb{<-} \hlnum{NA}
\hlstd{bird_raster}\hlkwb{<-} \hlkwd{staRVe_predict}\hlstd{(bird_compois,raster_to_pred)}
\end{alltt}
\end{kframe}
\end{knitrout}

\subsection{Haddock Survey}

To showcase the flexibility of our model we analyze a fairly complex dataset from a long running scientific survey on the Scotian shelf.
The survey spans the years of 1982 to 2016 with between 114 to 235 observations each year, except for one year with 30 observation, for a total of 6,259 observations across 35 years.
Each observation represents a scientific survey vessel fishing set along a spatial transect, from which we take the total biomass of haddock that is caught along that transect.
The transect locations are selected each year by a stratified random sampling scheme so the locations are not repeated from one year to the next.
We fit a tweedie model to the data then use the fitted model to forecast a predicted map of relative density for 2017.
We also include maps of predicted density for the years 2014-2016 to provide the recent context of the fishery.
These maps could then be used to help determine a management strategy for the haddock fishery.

The tweedie response distribution provides a continuous probability density function for non-zero observations and a point mass at zero to account for the zero-inflation typical of fisheries data \citep{Foster2013::PoissonGammaModel}.
We include the area swept by each fishing set as an offset term to account for differences in fishing effort between different observations.
% We also include quadratic effect of depth as the sole environmental covariate, since it can be reliably projected into the near future unlike other environmental factors such as sea bottom temperature.

\section{Results}

\subsection{Simulations}
\label{ssec::simResults}

Our brief simulation study shows that our model provides useful predictions with generally good point estimates and close to nominal 95\% confidence intervals.
As would be expected, forecast predictions are much more inaccurate than within-sample years but the confidence intervals are still close to nominal.
We interpret the plots of the simulations below to give an idea how the model parameters affect the model behaviour.

\paragraph{Spatial simulations with no observation error}

\begin{figure}
  \centering
  \includegraphics[width=\textwidth]{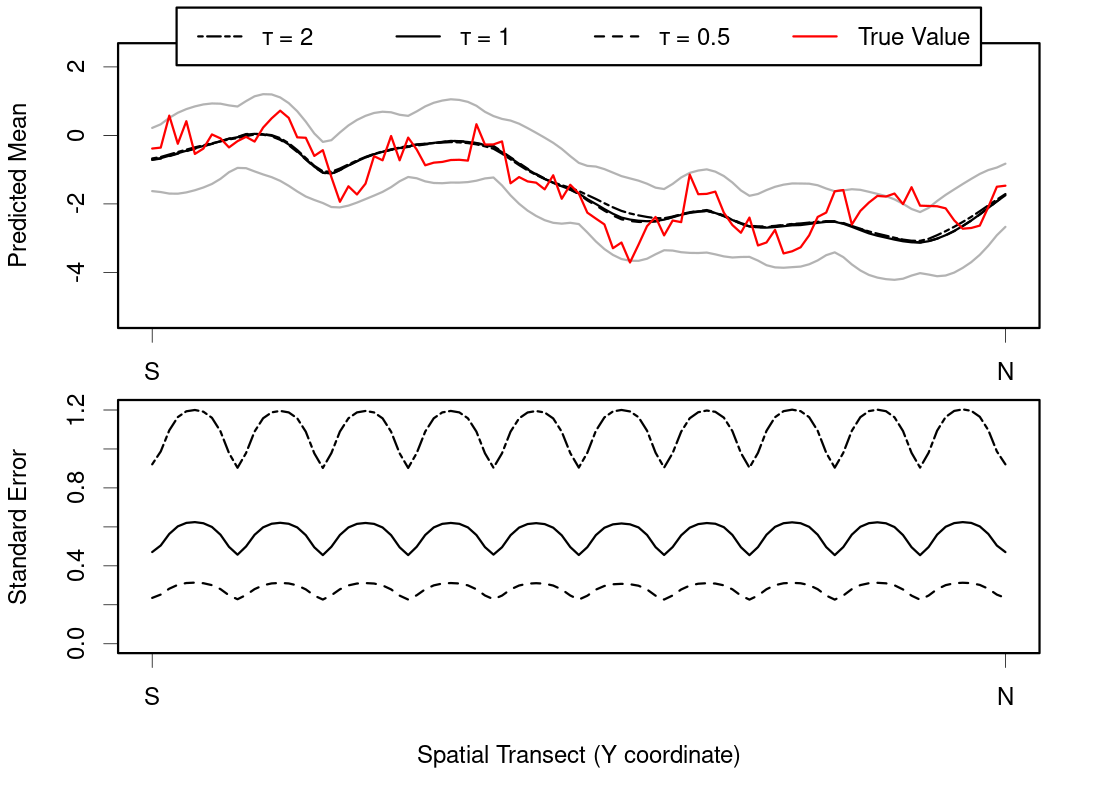}
  \caption{Simulated random effects (red) along the transect $(0.367,y)$ for $y\in[0,1]$.
    The spatial variance used to simulate the random effects is $\tau=1$.
    Predictions (top, black) and standard errors (bottom) for the random effects are obtained using different values of $\tau$.
    A 95\% confidence band (grey) is shown around the predicted mean when predicting under the same value of $\tau$ used to simulate the random effects.}
  \label{fig::spatialSim}
\end{figure}

Simulations and model predictions are shown in figure \ref{fig::spatialSim}.
The upper panel shows the predicted mean for the random effects with a 95\% confidence band around the predictions made using $\tau=1$, and the bottom panel shows the prediction standard error.
The mean predictions are nearly identical when using $\tau=0.5$, $\tau=1$, or $\tau=2$.
The bottom panel shows that the scale of the standard errors is directly proportional to the spatial standard deviation parameter.
For example the 95\% confidence band under $\tau=2$ is twice as wide as the 95\% confidence band using $\tau=1$, and four times as wide as the 95\% confidence band using $\tau=0.5$.

\paragraph{Spatio-temporal simulations with Gaussian response}

\begin{figure}
  \centering
  \includegraphics[width=0.9\textwidth]{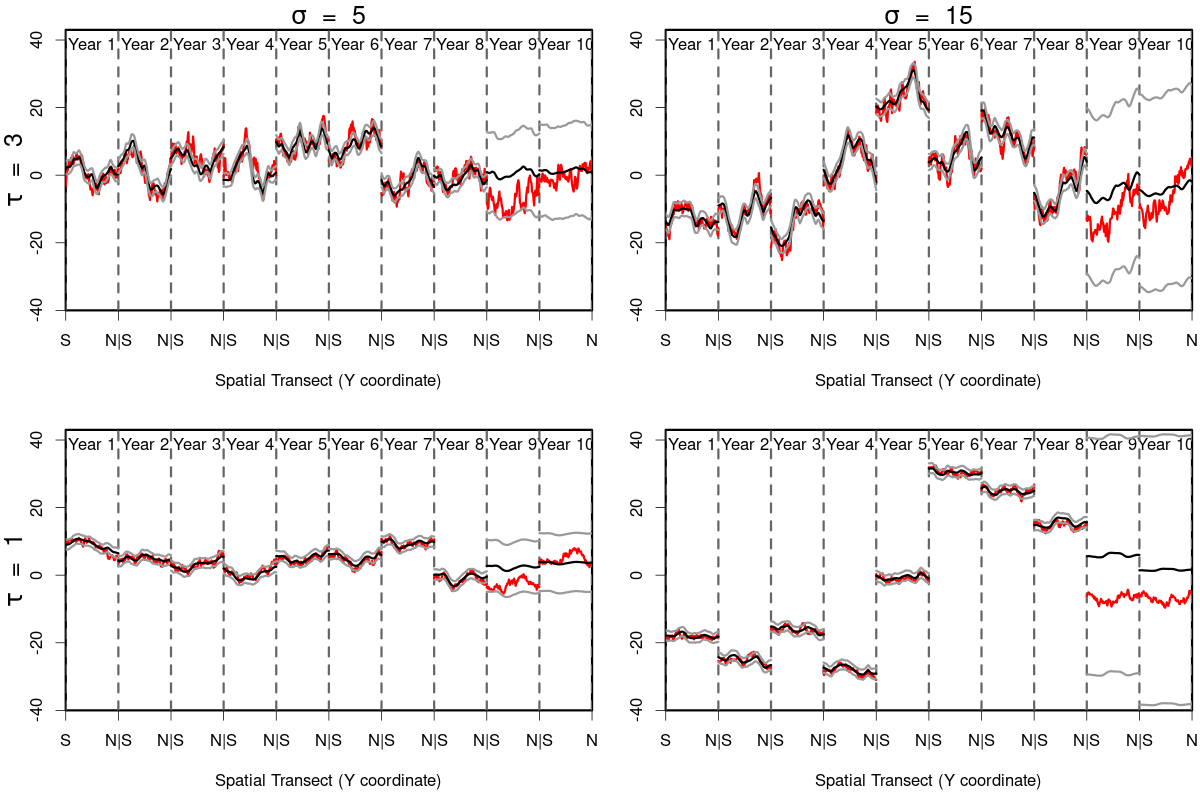}
  \caption{10 years of simulated (red) and predicted (black) random effects along the transect $(0.367,y)$ for $y\in[0,1]$ in a spatio-temporal model with a Gaussian response distribution. We have a 2 x 2 design with low and high spatial variance (bottom and top row, respectively) and low and high temporal variance (left and right column, respectively). The last two panels in each plot correspond to years 9 and 10. The fitted model did not include data from these years, so come as forecast predictions from the fitted model.}
  \label{fig::stObsSim}
\end{figure}

Simulations and fitted model predictions are shown in figure \ref{fig::stObsSim}.
The spatial and temporal variance parameters clearly affect the right components of the model behaviour.
A larger spatial variance corresponds to larger peaks and valleys within each year while a larger temporal variance exhibits bigger jumps in the overall mean between years.

\paragraph{Spatio-temporal simulations with Poisson response}

\begin{figure}
  \centering
  \includegraphics[width=0.9\textwidth]{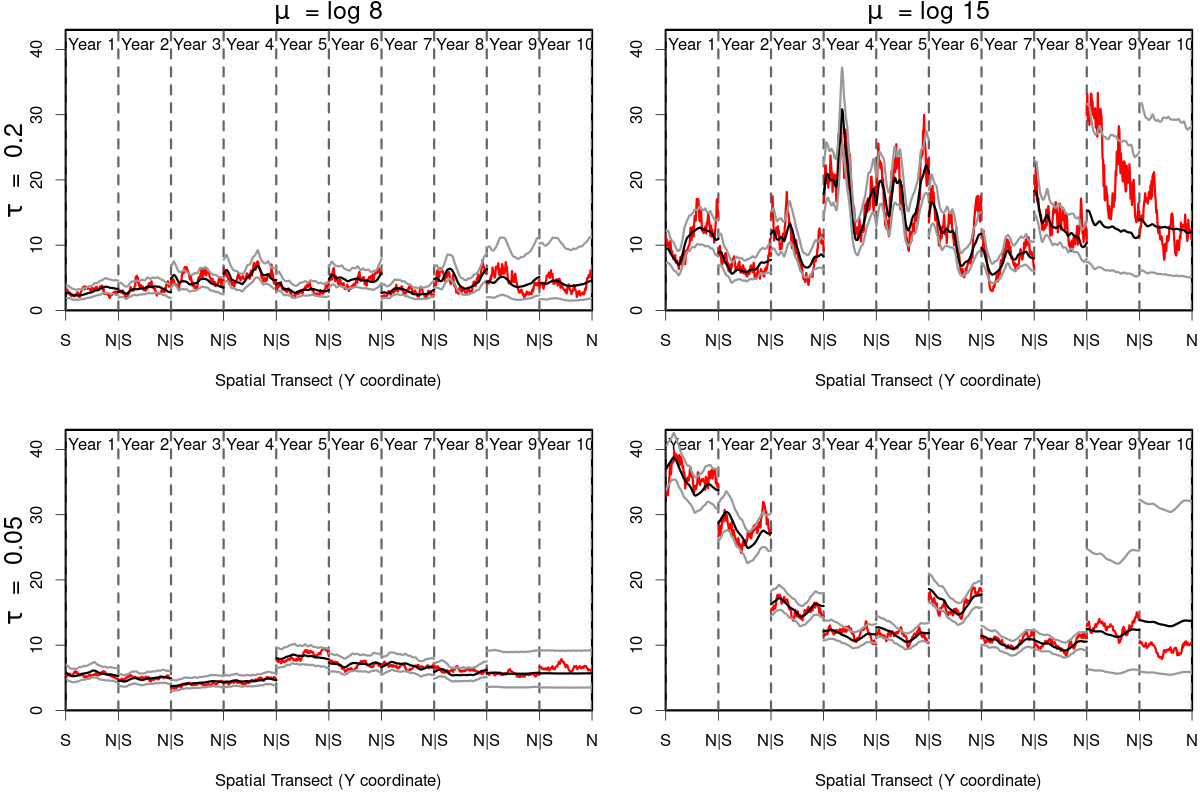}
  \caption{10 years of simulated (red) and predicted (black) response mean along the transect $(0.367,y)$ for $y\in[0,1]$ in a spatio-temporal model with a Poisson response distribution. We have a 2 x 2 design with low and high spatial variance (bottom and top row, respectively) and low and high average intensity (left and right column, respectively).}
  \label{fig::stPoisSim}
\end{figure}

Simulations and fitted model predictions are shown in figure \ref{fig::stPoisSim}.
The spatial variance acts the same as in the Gaussian case.
A larger average intensity exhibits a larger mean and increased overall variability, behaviour which is due to the log link function used.

\subsection{Carolina Wren Survey}

\begin{figure}
 \centering
 \includegraphics[width=0.6\textwidth]{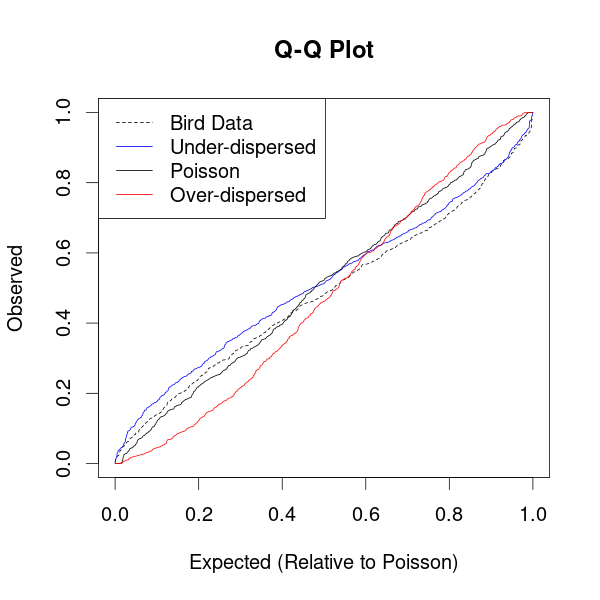}
 \caption{Q-Q plot comparing residual patterns for data sources with possible under- or over-dispersion. The observed dataset has residuals consistent with under-dispersion.}
 \label{fig::birdQQplot}
\end{figure}

The first attempt Poisson model successfully converged to a local maximum of the log-likelihood function, so we proceed to model validation.
The Q-Q plot of the Carolina wren residuals is shown in Figure \ref{fig::birdQQplot} alongside the residuals for simulated Poisson data, over-dispersed negative binomial data, and under-dispersed Conway-Maxwell-Poisson data.
The Q-Q plot of the Carolina wren residuals most closely resembles the Q-Q plot of the under-dispersed Conway-Mexwell-Poisson residuals, which suggests that the Carolina wren data exhibits under-dispersion relative to the Poisson model.
We refit the data using a Conway-Maxwell-Poisson distribution which also successfully converges.
The Conway-Maxwell-Poisson distribution is under-dispersed when the dispersion parameter is less than 1, over-dispersed when it is greater than 1, and becomes a (equidispersed) Poisson distribution when it is equal to 1.
The dispersion parameter is estimated to be \code{0.7919} with a standard error of \code{0.101} which confirms our suspicion of under-disperson.
Estimates of the other model parameters as well as spatio-temporal maps of the predicted mean intensity (with standard errors in a separate figure) are given in appendix \ref{app::birdDetails}.

\paragraph{Interpretation}

Missouri lies at the north-eastern edge of the Carolina wren's habitat range, and our analysis here seems to agree.
Figure \ref{fig::FullBirdPred} shows two population centers that are fairly stable throughout the years studed.
The first group lives in the south-eastern corner of the state, with the Mississippi River to the east and Mark Twain National Forest to the north-west of the population center.
The second population is present in the south-west corner of the state, on the northern edge of the Ozarks.
The predictions for this second population seem to be driven by observations at a single survey location with high counts.
Given the observations at this particular location are fairly unique, it would be worth investigating if the wren population is actually abundant at this location or if we are observing higher counts due to a particularly skilled bird watcher.
A log of the bird watcher names could identify if the same person ran the survey at that location each year, while a second independent survey could be used to try and validate the results from this dataset.

The model predictions show some interesting dynamics in the first eleven years.
From 1994 to 1997 the wren population is not very abundant, but seems to move south from around the St.~Louis area to the south-eastern corner of the state.
Then the population flourishes, and the sub-population north of the Ozarks appears as well.
But disaster strikes!
Between the survey in 2000 and the one in 2001, the Carolina wren population seems to completely vanish.
By 2006 the population has recovered and stays fairly stable until 2011.
Within these years, the south-eastern population and the Ozark population are not completely distinct groups as they are connected by a moderately abundant band across the Mark Twain National Forest.
In 2011 the population contracts spatially and wrens are seen in very low number everywhere except the south-eastern population, which is concentrated in a smaller geographic area than seen in the any of the previous years.
From 2011 to 2014 the Ozark population recovers once more and the south-eastern population shifts slightly northwards by about 50 kilometres, from the Sikeston area to the Cape Girardeau area.

A few questions arise from the above observations.
What caused the sudden population crash in 2001?
Was the decline in 2011 due to a similar cause?
What caused the subsequent population recoveries?
While our fitted model may not be able to answer these questions, it does provide a useful visualisation of the data that prompts us to ask these questions in the first place.
If we had included environmental covariates such as ground cover (forest, urban, farmland, etc.) or rainfall in the analysis, we might be able to explore the causes of the above behaviours more closely.

\subsection{Fish Survey}

Estimates of the model parameters as well as spatio-temporal maps of predicted density in the years 2014 to 2017 are given in appendix \ref{app::haddockDetails}.
First we look at the predictions in the non-forecast years, 2014 to 2017.
The maps show three discrete populations.
The most prominent is a large population centred just south of the southwestern tip of Nova Scotia that spreads north along the coastline and into the Bay of Fundy in the later years.
There seems to be a large population at the very southeastern edge of the study area in 2014, but since it is on the edge of the study area we don't feel it appropriate to interpret that prediction with this particular set of data.
The third population is directly south of Cape Breton and just west of Sable Island, the small sliver of land seen in the open ocean.
This population is a fair bit smaller than the first population both in geographical area occupied and in peak density, but the population seems to be fairly stable during the three years.
The area between the first and third populations is not completely uninhabitated by haddock, so while these two populations do seem to be distinct we might still expect some movement of individuals between the two populations.
Of course it is impossible to say for sure if this individual movement is occurring and an animal movement study or a capture-recapture study would be more appropriate to answer that question.
A genetic study would also show if these two populations are indeed distinct.

The forecast into the next year (2017) looks nearly identical to the map from 2016, but this is expected since the model uses an AR(1) structure and the AR(1) correlation was estimated to be very high at +0.95.
Arguably the forecasted standard errors are the more interesting part of the projections since they allow use to quantify how sure we are of the predictions.
The standard errors for the historically (at least from 2014-2016) uninhabited areas are very close to zero, so the model is fairly confident that new populations will not spontaneously arise from nowhere.
The standard errors for the population centres are higher, so while the model projects the population centres to be fairly stable it does not rule out the possibility of fairly dramatic changes in density in those areas.

Disregarding any insights we would gain from looking further back than 2014, we identify the emerging hotspot in the Bay of Fundy as an area of particular management interest.
The two main populations discussed above have a historical record which shows they are fairly stable and so management of those areas probably does not need to be altered much.
The new hotspot in the Bay of Fundy doesn't have the same longevity and deserves more thought.
If conservation is the goal, would it be worth spending the resources to designate that area as a temporary fisheries closure to try and protect the new population? Or will the haddock have moved back to the main population in a year and those resources would be better spent elsewhere?
If fishing is the goal, is the allure of a high-risk high-reward area close to the shoreline enough to drag part of the fishing fleet away from the more stable areas that are farther out in the ocean?
No statistical model can answer these questions, but the visualisations certainly provide a useful starting point for identifying areas of interest and directions for future studies.

\section{Discussion}
\label{sec::discussion}

The emergence and importance of citizen science in ecology, the ubiquity of atmospheric raster data from satellite images, and the desire to understand complex geographical processes have created a huge collection of spatio-temporally referenced point data.
While these datasets can be quite complex, they are often conceptually similar and are too numerous for each researcher to have to implement their own model from scratch.
We have introduced a new modelling framework that can be used for common ecological survey data, expanding the toolbox available to researchers.

The structure of our modelling framework provides access to simulation-based methods for model interpretation and validation which is otherwise quite difficult for spatio-temporal data.
We showed one method for checking the response distribution using simulation methods but the fitted model could also have other problems such as collinearity in covariates, a mis-specified structure for the temporal random effects, an incorrect spatial covariance function, and more.
The relevant literature is rather undeveloped and there are not many common practices, but the method we use for the response distribution could potentially be modified to inspect other parts of the model.
The general procedure would be simulating datasets from multiple alternative models including the fitted model, and comparing the residuals from the original dataset to those of the simulated datasets.
Future research could focus on which alternative models to use and how to compare between them in order to check for validity of the temporal and spatial structures assumed in the model.

We end with a discussion of data sources and applications that could benefit from our modelling framework.

\paragraph{eBird}

A real-time, online checklist program, eBird has revolutionised the way that the birding community reports and accesses information about birds.
eBird provides rich data with basic information on bird abundance and distribution at a variety of spatial and temporal scales.
eBird is one of the largest and fastest growing biodiversity data resources in existence.

This data enables us to look at population changes over space and time.
If we consider data from Nova Scotia, Canada, there are about 25,000 checklists provided by about 600 observers on a yearly basis with about 350 species reported annually.
A particular question of interest relates to changes in abundance of particular species over time and space in response to climate change.
For species such as boreal chickadee and Bicknell’s thrush where Nova Scotia is at the southern limit of their range, anecdotal evidence shows these species much less abundant in southern Nova Scotia in recent years.
In addition Bicknell’s thrush is a bird of the highlands and boreal forest.
With the models proposed in this paper we can answer important questions about their population trends temporally and spatially.

\paragraph{Disease Mapping}

Disease mapping has its own set of complications -- preferential sampling and privacy concerns among them -- but the basic scenario is similar to the Carolina wren analysis.
We refer to \citet{Giorgi::geostatisticalDiseaseMapping} for an introduction to spatio-temporal models for disease prevalence.
The model they explore is a binomial GLMM which can be fit using the \pkg{staRVe} package.

Many disease prevalence datasets must comply with privacy regulations, and thus are typically reported as areal data.
However following \citet{Giorgi::geostatisticalDiseaseMapping} these areal data can sometimes be treated as point-referenced data.
Their example samples individual villages and treats each village as a single point.
In this case the size of the village is small compared to the distance between villages, which allows them to treat the observations as point data.

In other cases where data is point-referenced but still restricted by privacy concerns the data are usually available to only a small set of specialists, perhaps containing no dedicated statisticians.
Our package could serve a vital role here in supplying an easy to use method with little advanced coding necessary.

\paragraph{Fisheries Science}

The original motivating example for our development of the \pkg{staRVe} package is the analysis of fisheries data in Atlantic Canada, so we add to some of the more general ideas introduced in our analysis of haddock.
Imagine we had collected data on lobster abundance off the coast of Maine and discovered a potential northward shift which could have huge economic consequences.
A large enough shift north or east, even in the range of 150km, could see the lobster stock cross international boundaries from Maine in the United States to New Brunswick and Nova Scotia in Canada, making the lobster unavailable to American fishers (but available to Canadian fishers).
Identifying the causes of these shifts can help determine if they are likely to be temporary or permanent.
In addition, reliable forecasts of lobster stock behaviour are invaluable in planning any stock management strategies, ecosystem protection, or economic decisions in response to these changes.

\section*{Acknowledgments}
\label{sec::acknowledgments}

This research was funded by a Canadian Statistical Sciences Institute Collaborative Research Team Project and the Ocean Frontier Institute. EL is also supported by a Vanier Canada Graduate Scholarship and Killam Predoctoral Scholarship.
We thank the Department of Fisheries and Oceans Canada for providing the survey data used in the haddock analysis.

\section*{Conflict of Interest Statement}

The authors have no conflicts of interest to disclose.

\section*{Author Contributions}

EL designed the model and package with input from CF and JMF. All authors contributed to writing the paper and approved it for publication.

\section*{Data Availability}

The data used in this paper are included as part of the \pkg{staRVe} package, which is available at the first author's GitHub page (\url{https://github.com/lawlerem/staRVe}).
The specific version of the package used for this manuscript is archived at doi:10.5281/zenodo.5548203.

\bibliographystyle{apalike}
\bibliography{bibliography}

\newpage
\appendix

\section{Figures and Tables}
\subsection{Parameter estimates and predictions for the Carolina wren}
\label{app::birdDetails}

\begin{table}[H]
\begin{knitrout}
\definecolor{shadecolor}{rgb}{0.969, 0.969, 0.969}\color{fgcolor}
\begin{tabular}{l|r|r}
\hline
  & Estimate & Standard Error\\
\hline
(log) global mean $\mu$ & 1.8659 & 0.5879\\
\hline
AR(1) correlation $\phi$ & 0.9250 & 0.0221\\
\hline
temporal std. dev. $\sigma$ & 0.7535 & 0.1907\\
\hline
spatial std. dev. $\tau$ & 0.4649 & 0.0681\\
\hline
spatial range $\rho$ & 72.9335 & 25.0282\\
\hline
CMP dispersion & 0.7919 & 0.1010\\
\hline
\end{tabular}

\end{knitrout}
  \caption{Parameter estimates for Conway-Maxwell-Poisson model fitted to the Carolina wren dataset}
\end{table}

\begin{figure}[H]
\begin{knitrout}
\definecolor{shadecolor}{rgb}{0.969, 0.969, 0.969}\color{fgcolor}
\includegraphics[width=\maxwidth]{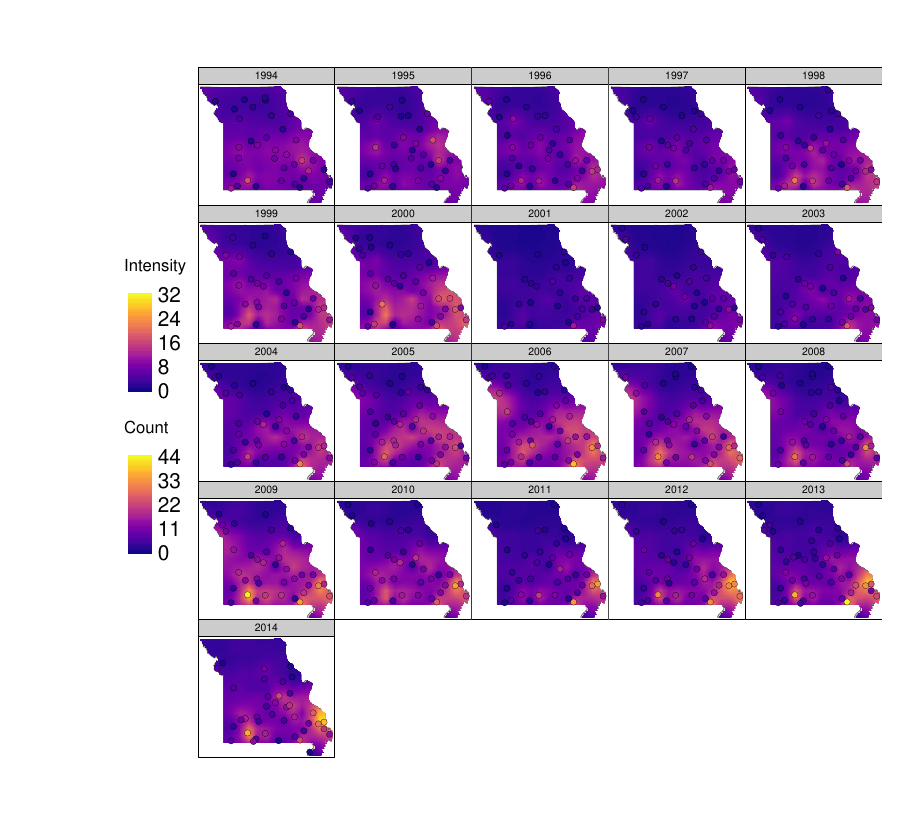} 

\end{knitrout}
	\caption{Model-based predictions of mean intensity for all years of the Carolina wren dataset.}
	\label{fig::FullBirdPred}
\end{figure}

\begin{figure}[H]
\begin{knitrout}
\definecolor{shadecolor}{rgb}{0.969, 0.969, 0.969}\color{fgcolor}
\includegraphics[width=\maxwidth]{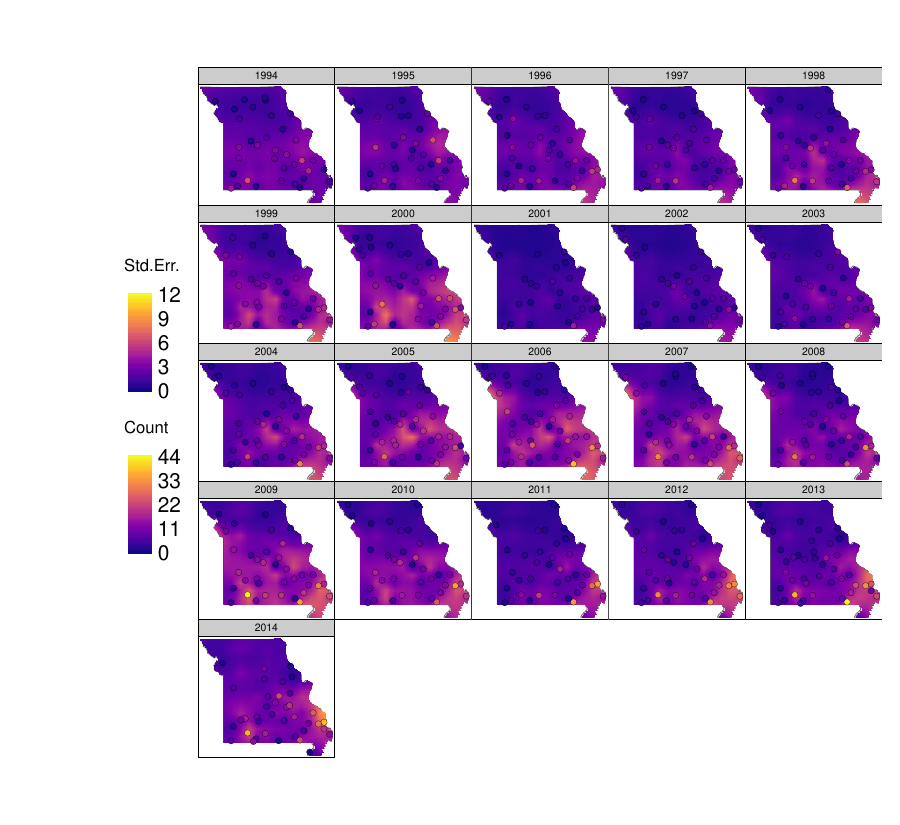} 

\end{knitrout}
	\caption{Model-based standard errors for predicted mean intensity for all years of the Carolina wren dataset.}
	\label{fig::FullBirdSE}
\end{figure}

~
\newpage
~

\subsection{Parameter estimates and predictions for the haddock survey}
\label{app::haddockDetails}

\begin{table}[H]
\begin{knitrout}
\definecolor{shadecolor}{rgb}{0.969, 0.969, 0.969}\color{fgcolor}
\begin{tabular}{l|r|r}
\hline
  & Estimate & Standard Error\\
\hline
(log) global mean $\mu$ & -4.8936 & 0.6844\\
\hline
AR(1) correlation $\phi$ & 0.9536 & 0.0071\\
\hline
temporal std. dev. $\sigma$ & 0.8200 & 0.1859\\
\hline
spatial std. dev. $\tau$ & 1.5037 & 0.0651\\
\hline
spatial range $\rho$ & 35.0967 & 2.7712\\
\hline
tweedie dispersion & 3.7144 & 0.1007\\
\hline
tweedie power & 1.5750 & 0.0065\\
\hline
\end{tabular}

\end{knitrout}
  \caption{Parameter estimates for tweedie model fitted to the haddock survey}
\end{table}

\begin{figure}[H]
\begin{knitrout}
\definecolor{shadecolor}{rgb}{0.969, 0.969, 0.969}\color{fgcolor}
\includegraphics[width=\maxwidth]{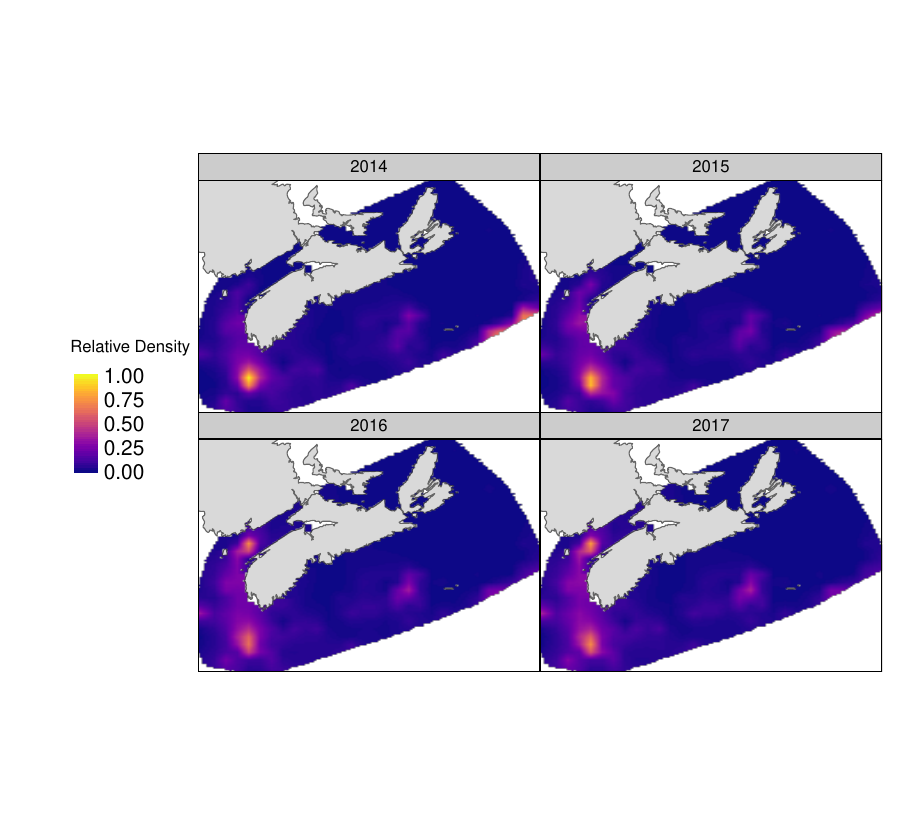} 

\end{knitrout}
  \caption{Model-based predictions of relative density for 2014-2017 of the haddock survey.}
  \label{fig::fullHaddockPred}
\end{figure}

\begin{figure}[H]
\begin{knitrout}
\definecolor{shadecolor}{rgb}{0.969, 0.969, 0.969}\color{fgcolor}
\includegraphics[width=\maxwidth]{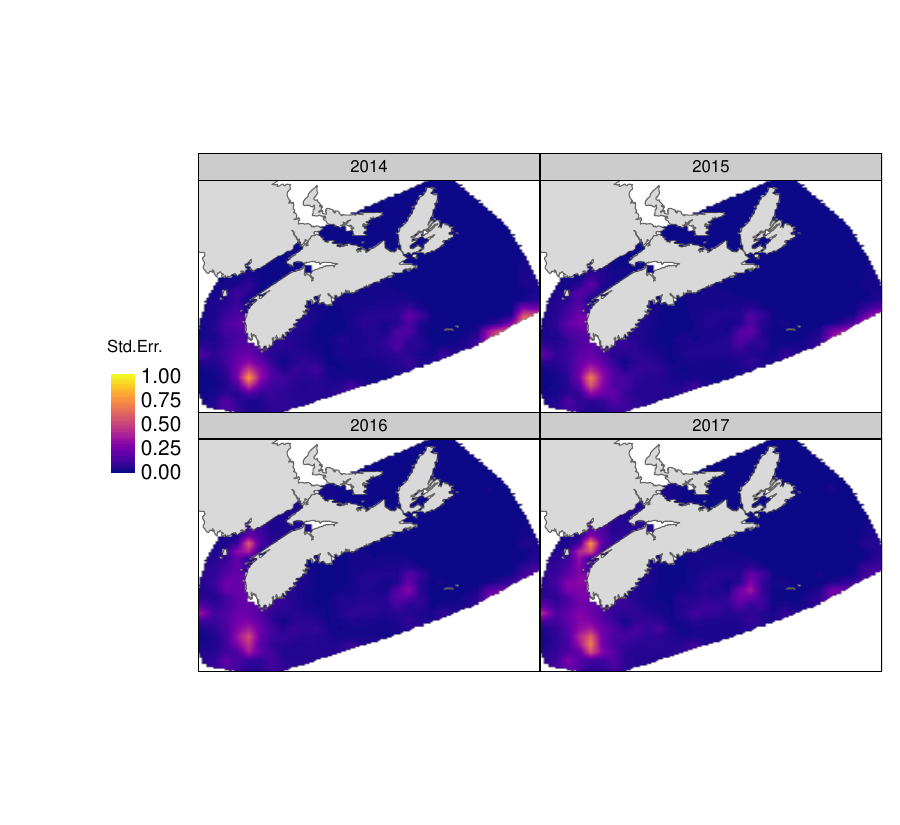} 

\end{knitrout}
  \caption{Model-based standard errors of predicted relative density for 2014-2017 of the haddock survey.}
  \label{fig::fullHaddockSE}
\end{figure}

~
\newpage
~

\section{More details on the NNGP log-likelihood}
\label{app::likelihoodDetails}

Here we detail the computation of the log-likelihood contribution $\ell_{w}\left(\Theta\right)$ for the spatio-temporal nearest-neighbour Gaussian process (NNGP), in particular the "spatial" level of the hierarchy.
We note that there are modelling choices that could have been made differently from the ones we have implemented and describe below.
We made our choices based on a combination of model simplicity, computational efficiency, and personal preference.
The original paper introducing the NNGP in a spatial context is \citet{Datta2016::nearestNeighbourGP}, and to learn more about the broader class of probabilistic graphical models we suggest \citet{Koller2009::ProbabilisticGraphicalModels}.

A Gaussian process is a (possibly infinite) collection of random variables, where any finite subset of that collection has a multivariate normal distribution.
In the spatial context, a Gaussian process is specified completely by a mean function and a covariance function.
Throughout this section we speak of the locations and the random effects at those locations interchangeably.
For inference we look at the joint distribution of the observed locations and any additional locations we want to include.

We start by defining the nearest neighbour Gaussian process in the purely spatial case.
Evaluating the joint density of a regular Gaussian process for even a moderately sized number of locations is computationally challenging because it requires computing the inverse of the covariance matrix.
Using basic rules of conditional probabilities, we can factor the joint density as a product of (univariate) conditional densities:
\[
  f\left(\mathbf{W}\left(\mathbf{s}_{1:N}\right)\right) =
    f_{1}\left(W\left(\mathbf{s}_{1}\right)\right)\cdot\prod_{i=1}^{N-1} f_{i+1}\left(W\left(\mathbf{s}_{i+1}\right)~\vert~\mathbf{W}\left(\mathbf{s}_{1:i}\right)\right)
\]
where we write $\mathbf{W}\left(\mathbf{s}_{1:i}\right)$ to mean random variables at the first $i$ location of the multivariate normal variable.
This isn't immediately useful, as the term $f_{N}\left(W\left(\mathbf{s}_{N}\right)~\vert~\mathbf{W}\left(\mathbf{s}_{1:(N-1)}\right)\right)$ is almost as computationally expensive to compute as the full joint density.

However, the nearest neighbour Gaussian process imposes a set of conditional independence constraints.
These constraints allows us to replace the conditioning set $\mathbf{W}\left(\mathbf{s}_{1:i}\right)$ with a much smaller one, which drastically reduces the computational burden of computing the densities.
To describe the conditional independence constraints we introduce a set of reference locations $\mathcal{S}_{per}$ and give them an order.
In our implementation the first location in the order is arbitrary, and subsequent locations in the order are taken to be whichever as yet unordered location is closest to any location that is already ordered.
A location $\mathbf{s}$ in $\mathcal{S}_{per}$ is conditionally independent of all other locations, given the $k$ geographically closest locations in $\mathcal{S}_{per}$ that came before it in the order.
A location $\mathbf{s}$ not in $\mathcal{S}_{per}$ is conditionally independent of all other locations, given the $k$ geographically closest locations in $\mathcal{S}_{per}$ regardless of the order.
We call these "$k$ closest locations" the parents of $\mathbf{s}$, or $\delta\mathbf{s}$ for short, and choose $k$ to be some small number e.g.,~$k=15$.

Given a set of observed locations and user-chosen $\mathcal{S}_{per}$, we encode the above conditional independence constraints in a directed acyclic graph (DAG) named $\mathcal{G}$.
The vertices of the graph are the set of all locations considered.
There is a directed edge from vertex $i$ to vertex $j$ iff $\mathbf{s}_{j}$ is a parent of $\mathbf{s}_{i}$.

With these conditional independence constraints, the joint density of locations $\mathbf{s}_{1:N}$ and reference locations $\mathbf{S}_{per}$ is
\[
  f\left(\mathbf{W}\left(\mathbf{s}_{1:N}\right),~\mathbf{W}\left(\mathcal{S}_{per}\right)\right) =
    \prod_{\mathbf{s}\in\mathcal{S}_{per}}f_{\mathbf{s}}\left(W\left(\mathbf{s}\right)~\vert~\mathbf{W}\left(\delta\mathbf{s}\right)\right)\cdot
    \prod_{i=1}^{N} f_{i}\left(W\left(\mathbf{s}_{i}\right)~\vert~\mathbf{W}\left(\delta\mathbf{s}_{i}\right)\right).
\]
Thus $\mathcal{G}$ also describes the series of conditional densities used to compute the joint density.

We generalise the above to our spatio-temporal setting by creating a graph $\mathcal{G}_{t}$ for each time point $t$ which is split in two parts: the persistent graph $\mathcal{G}_{per}$ corresponding to the reference locations $\mathcal{S}_{per}$, and the transient graphs $\mathcal{G}_{trans,t}$ encoding the parents $\delta\mathbf{s}_{i}$ of the observed locations at time point $t$.
Since the reference locations do not change from year to year, the persistent graph $\mathcal{G}_{per}$ and thus the conditional independence constraints and density factorization are the same from year to year.
Combined with property that Gaussian conditional variances do not depend on the realised value of the random variables conditioned upon, this allows us to cache most of the computations used in computing the likelihood for the persistent graph.
The transient graphs $\mathcal{G}_{trans,t}$ can change from year to year since different locations may be observed, but a location $\mathbf{s}$ will have the same parents every year.

We split the computation of $\ell_{w}\left(\Theta\right)$ into contributions $\ell_{per}\left(\Theta\right)$ from the reference locations and $\ell_{tran}\left(\Theta\right)$ from the observed locations.

\subsection{Persistent graph log-likelihood}
\label{sapp::persistentLikelihood}

Here we decribe the conditional distributions used for the reference locations and persistent graph.
We recall our notation that $\mathbf{W}_{t}\left(\mathbf{s}\right)$ is the random effect at time point $t$ and location $\mathbf{s}$.
We also use $\mathbf{W}_{t}\left(\delta\mathbf{s}\right)$ to be joint random effects at the locations in $\delta\mathbf{s}$.
As a reminder the "temporal" level in the model hierarchy defines a set of purely temporal random effects defined by
\[
  \bm{\epsilon} \sim
    \text{N}\left(\mu,~\frac{\sigma^{2}}{1-\phi^{2}}\cdot \Sigma\right)
    ~~\text{where}~~\Sigma_{ij} = \phi^{\norm{i-j}}.
\]
An equivalent formulation that will be useful later is
\begin{align*}
  \epsilon_{t} ~\vert \epsilon_{t-1} &= \phi\cdot\epsilon_{t-1} + \nu \tag{$\nu\sim N\left(0,\sigma^{2}\right)$} \\
  \epsilon_{1} &= \mu + \nu \tag{$\nu\sim N\left(0,\frac{\sigma^{2}}{1-\phi^{2}}\right)$}
\end{align*}

For presentation purposes we use the exponential covariance function
\[
  \mathcal{C}\left(\mathbf{s}_{1},\mathbf{s}_{2}\right)
    = \tau^{2}\rho
      \Exp{-\frac{  \norm{\mathbf{s}_{1}-\mathbf{s}_{2}}  }{  \bar{d}  }
        \cdot \frac{1}{\rho} },
\]
but any Mat\'ern covariance function can be used.

At the initial time point $t=1$ we assume a constant marginal mean equal to $\epsilon_{1}$.
The conditional distribution for $W_{1}\left(\mathbf{s}_{i}\right)$ is
\[
	W_{1}\left(\mathbf{s}_{i}\right)~\bigr\vert~ \mathbf{w}_{1}\left(\delta\mathbf{s}_{i}\right),~\bm{\epsilon}
		\sim N\left(
				\epsilon_{1}+\mathbf{c}^{\top}\Sigma^{-1} \left(\mathbf{w}_{1}\left(\delta\mathbf{s}_{i}\right)-\epsilon_{1}\right),~
				\tau^{2}\rho - \mathbf{c}^{\top}\Sigma^{-1}\mathbf{c}
			\right)
\]
where $\delta\mathbf{s}_{i}$ is the set of parents of $\mathbf{s}_{i}$ in $\mathcal{G}_{per}$, $\mathbf{c}^{\top}$ is the cross-covariance vector between $W_{1}\left(\mathbf{s}_{i}\right)$ and $\mathbf{W}_{1}\left(\delta\mathbf{s}_{i}\right)$,
 	$\Sigma$ is the covariance matrix of $\mathbf{W}_{1}\left(\delta\mathbf{s}_{i}\right)$, and $\tau^{2}\rho$ is the marginal spatial variance.
In the spatial statistics literature, the above conditional mean and variance are known as the kriging mean and variance.
The likelihood for the first $k$ random effects in the persistent graph is computed through their joint distribution
\[
  \mathbf{W}_{1}\left(\mathbf{s}_{1:k}\right) \sim
    N\left(\epsilon_{1},~\Sigma\right),
\]
where here $\Sigma$ is the covariance matrix of $\mathbf{W}_{1}\left(\mathbf{s}_{1:k}\right)$.

For times $t>1$ we define the marginal mean of $\mathbf{W}_{t}\left(\mathbf{s}\right)$ to be $\mathbf{m}_{t}\left(\mathbf{s}\right)=\phi\cdot \left(\mathbf{w}_{t-1}\left(\mathbf{s}\right)-\epsilon_{t-1}\right)+\epsilon_{t}$.
This mean function gives a marginal AR(1) process to each location, which we verify by showing the (marginal) mean for $\mathbf{W}_{t}\left(\mathbf{s}\right)~\vert~\mathbf{w}_{t-1}\left(\mathbf{s}\right)$ is $\phi\cdot \mathbf{w}_{t-1}\left(\mathbf{s}\right)$:
\begin{align*}
  \E{\mathbf{W}_{t}\left(\mathbf{s}\right)~\vert~\mathbf{w}_{t-1}\left(\mathbf{s}\right),~\bm{\epsilon}}
    &= \phi\cdot\left(\mathbf{w}_{t-1}\left(\mathbf{s}\right)-\epsilon_{t-1}\right)+\epsilon_{t} \\
    &= \phi\cdot \mathbf{w}_{t-1}\left(\mathbf{s}\right) + \left(\epsilon_{t}-\phi\cdot\epsilon_{t-1}\right) \\
\intertext{The distribution for $\epsilon_{t}-\phi\cdot\epsilon_{t-1}$ is a zero-centered (symmetric) normal distibution, so marginalizing out $\bm{\epsilon}$ gives}
  \E{\mathbf{W}_{t}\left(\mathbf{s}\right)~\vert~\mathbf{w}_{t-1}\left(\mathbf{s}\right)}
    &= \phi\cdot\mathbf{w}_{t-1}\left(\mathbf{s}\right).
\end{align*}

Using the above mean function $\mathbf{m}_{t}\left(\mathbf{s}\right)=\phi\cdot \left(\mathbf{w}_{t-1}\left(\mathbf{s}\right)-\epsilon_{t-1}\right)+\epsilon_{t}$, the conditional distribution for $W_{t}\left(\mathbf{s}_{i}\right)$ is then
  \[
  	\mathbf{W}_{t}\left(\mathbf{s}_{i}\right)~\bigr\vert~ \mathbf{w}_{t-1}\left(\delta\mathbf{s}_{i}\right),~\bm{\epsilon}
  		\sim N\left(
  				\mathbf{m}_{t}\left(\mathbf{s}_{i}\right)+\mathbf{c}^{\top}\Sigma^{-1} \left(\mathbf{w}_{t}\left(\delta\mathbf{s}_{i}\right)-\mathbf{m}_{t}\left(\delta\mathbf{s}_{i}\right)\right),~
  				\tau^{2}\rho - \mathbf{c}^{\top}\Sigma^{-1}\mathbf{c}
  			\right).
  \]
As in the case for $t=0$, the likelihood for the first $k$ random effects in the persistent graph is computed through their joint distribution
  \[
    \mathbf{W}_{t}\left(\mathbf{s}_{1:k}\right) \sim
      N\left( \phi\cdot\left(\mathbf{w}_{t-1}\left(\mathbf{s}_{1:k}\right)-\epsilon_{t-1}\right) + \epsilon_{t},~\Sigma\right).
  \]

There is a subtle identifiability issue related to the first $k$ random effects in the persistent graph of each year, which we solve in an admittedly strange-looking fashion.
The identifiability issue arises because the spatial range parameter $\rho$ is unidentifiable and contributes to the marginal spatial variance $\tau^{2}\rho$.
Both the marginal spatial variance and the temporal variance (by virture of $\epsilon_{t}$ appearing in the mean of the spatial random effects) control the total variance of the first $k$ random effects, so the unidentifiability of $\rho$ creeps in to cause the spatial variance to be unidentifiable.
To deal with these crossed wires we temporarily use a different value of $\rho$ for the first $k$ random effects, choosing the value that sets the marginal variance equal to the average conditional spatial variance of the remaining random effects in the persistent graph.
This forces the spatial component of the variance for the first $k$ random effects to be similar in magnitude to the spatial component of the variance for the remaining random effects, and results in the temporal variance being identifiable.
Moreover, since the value of $\rho$ does not affect the conditional distribution this strange looking solution is technically sound.

\subsection{Transient node log-likelihood}
\label{sapp::transientLikelihood}

The conditional distributions for the observation locations and transient graphs require only a small change from the distributions in Section \ref{sapp::persistentLikelihood}.
Since a location $\mathbf{s}$ observed at time point $t$ may not have been observed at time point $t-1$, we don't have a random effect $w_{t-1}\left(\mathbf{s}\right)$ explicitly modelled to use in
  \(
    \mathbf{m}_{t}\left(\mathbf{s}\right) = \phi\cdot \left(w_{t-1}\left(\mathbf{s}\right)-\epsilon_{t-1}\right)+\epsilon_{t}.
  \)
We replace $w_{t-1}\left(\mathbf{s}\right)$ with the predictor
  \[
    \tilde{w}_{t-1}\left(\mathbf{s}\right) =
      \frac{\mathbf{c}^{\top}\Sigma^{-1}\mathbf{w}_{t-1}\left(\delta\mathbf{s}_{i}\right)}
				{\mathbf{c}^{\top}\Sigma^{-1}\mathbf{1}}
  \]
  where $\mathbf{1}$ is a vector of ones.
This prediction is the best linear unbiased predictor of the mean of the process $\mathbf{W}_{t-1}\left(\delta\mathbf{s}\right)$ in an ordinary kriging framework \citep{Cressie1990::originsKriging}.
Since this is a weighted average of $\mathbf{w}_{t-1}\left(\delta\mathbf{s}\right)$, this is a local estimator of the mean.
The conditional distribution is then
  \[
  	W_{t}\left(\mathbf{s}\right)~\bigr\vert~ \mathbf{w}_{t}\left(\delta\mathbf{s}\right),~\bm{\epsilon}
  		\sim N\left(
  				\tilde{\mathbf{m}}_{t}\left(\mathbf{s}_{i}\right)+\mathbf{c}^{\top}\Sigma^{-1} \left(\mathbf{w}_{t}\left(\delta\mathbf{s}\right)-\mathbf{m}_{t}\left(\delta\mathbf{s}\right)\right),~
  				\sigma_{s}^{2} - \mathbf{c}^{\top}\Sigma^{-1}\mathbf{c}
  			\right)
  \]
  where
  \(
    \tilde{\mathbf{m}}_{t}\left(\mathbf{s}\right) = \phi\cdot \left(\tilde{w}_{t-1}\left(\mathbf{s}\right) - \epsilon_{t-1}\right) + \epsilon_{t}.
  \)

\newpage
\section{More details on predictions}
\label{app::predictionDetails}

We expand on our description of our prediction procedure given in the main text.
Our aim is to produce predictions for the random effect $W_{t}\left(\mathbf{s}\right)$, the linear predictor on the link scale $L_{t}\left(\mathbf{s}\right)$, and the response mean $\mu_{t}\left(\mathbf{s}\right)$ at some unobserved spatial location $\mathbf{s}$ and time point $t$.
Since there is uncertainty involved in the inference of the random effects and fixed parameters $\Theta = \left\{\theta,~\bm{\beta},~\phi,~\mu,~\sigma^{2},~\tau^{2}\right\}$ we also want to propogate this uncertainty to our prediction standard errors.

\subsection{The random effects}
\label{sapp:randomEffectPrediction}

The predictive distribution for the new random effect $W_{t}\left(\mathbf{s}_{i}\right)$ is the same as the conditional distributions we used for the observation locations in Section \ref{sapp::transientLikelihood}, namely
  \[
  	W_{t}\left(\mathbf{s}\right)~\bigr\vert~ \mathbf{w}_{t}\left(\delta\mathbf{s}\right),~\bm{\epsilon}
  		\sim N\left(
  				\tilde{\mathbf{m}}_{t}\left(\mathbf{s}_{i}\right)+\mathbf{c}^{\top}\Sigma^{-1} \left(\mathbf{w}_{t}\left(\delta\mathbf{s}\right)-\mathbf{m}_{t}\left(\delta\mathbf{s}\right)\right),~
  				\sigma_{s}^{2} - \mathbf{c}^{\top}\Sigma^{-1}\mathbf{c}
  			\right)
  \]
  where
  \(
    \tilde{\mathbf{m}}_{t}\left(\mathbf{s}\right) = \phi\cdot \left(\tilde{w}_{t-1}\left(\mathbf{s}\right) - \epsilon_{t-1}\right) + \epsilon_{t}.
  \)
To the joint likelihood $\ell_{J}\left(\Theta\right)$ we add a likelihood contribution $\ell_{p}\left(\Theta\right)$ using the above conditional distribution.
The predicted value for $W_{t}\left(\mathbf{s}\right)$ is then taken to be the mode of $\ell_{J}\left(\hat{\Theta}\right)+\ell_{p}\left(\hat{\Theta}\right)$ as a function of $w_{t}\left(\mathbf{s}\right)$, and the standard errors are computed from the hessian evaluated at the mode.
This is essentially a plug-in estimate using the maximum likelihood estimates $\hat{\Theta}$, but since we take the predictions directly from the likelihood function all uncertainty in parameter estimation is propogated to the predictions.

\subsection{The linear predictor}
\label{sapp::linearMeanPrediction}

Having estimates and standard errors for $\hat{W}_{t}\left(\mathbf{s}_{i}\right)$ and $\hat{\bm{\beta}}$ and having a set of prediction covariates $\mathbf{X}$, we can easily find the predictions and standard errors for the linear predictor
	\begin{align*}
		\hat{L}_{t}\left(\mathbf{s}\right) &=
			\mathbf{X}\hat{\bm{\beta}} + \hat{w}_{t}\left(\mathbf{s}_{i}\right) \\
		\Var{\hat{L}_{t}\left(\mathbf{s}\right)} &=
			\mathbf{X}\Var{\hat{\bm{\beta}}}\mathbf{X}
					+ \Var{\hat{w}_{t}\left(\mathbf{s}_{i}\right)}
	\end{align*}
In the variance calculation we are assuming the estimator $\hat{\bm{\beta}}$ is uncorrelated with $\hat{w}_{t}\left(\mathbf{s}\right)$.

\subsection{The response mean}
\label{sapp::responePrediction}

Our last objective is to predict the response mean
  \(
    \hat{\mu}_{t}\left(\mathbf{s}\right) = g^{-1}\left[\hat{L}_{t}\left(\mathbf{s}\right)\right].
  \)
Letting $h\left(x\right) = (g^{-1})''\left(x\right)$ be the second derivative of $g^{-1}\left(x\right)$,
	we apply a second-order Taylor approximation for the mean and variance of $g^{-1}\left[\hat{L}_{t}\left(\mathbf{s}\right)\right]$:
	\begin{align*}
		\E{g^{-1}\left(\hat{L}_{t}\left(\mathbf{s}\right)\right)}
			&\approx g^{-1}\left(\hat{L}_{t}\left(\mathbf{s}\right)\right)
				+ \frac{1}{2}\cdot h\left(\hat{L}_{t}\left(\mathbf{s}\right)\right) \\
		\Var{g^{-1}\left(\hat{L}_{t}\left(\mathbf{s}\right)\right)}
			&\approx g^{-1}\left(\hat{L}_{t}\left(\mathbf{s}\right)\right)^{2}\cdot \Var{\hat{L}_{t}\left(\mathbf{s}\right)}
				+ \frac{1}{2}\cdot h\left(\hat{L}_{t}\left(\mathbf{s}\right)\right)^{2}\cdot
					\Var{\hat{L}_{t}\left(\mathbf{s}\right)}^{2}
	\end{align*}

\subsection{Prediction intervals for new observations}

Keeping with accepted practice for generalized linear models we do not attempt to create prediction intervals for new observations.
While computing closed form prediction intervals is relatively straightforward for a Gaussian response distribution such intervals either may not make sense, in the case of the bernoulli distribution, or may not have a closed form, as in the case of a Poisson distribution.
However observations are assumed conditionally independent of all other observations given the covariates and spatio-temporal random effects.
The predicted response mean and estimated response distribution parameters are all that are needed to produce probability statements for any particular new observation, including prediction intervals, which could be found either from the density function of the response distribution or through simulations of new observations conditional on the fitted random effects.

\end{document}